\documentclass[preprint]{vgtc}                          % final (conference style)
\graphicspath{{figures/}{pictures/}{images/}{./}} % where to search for the images

\usepackage{times}                     % we use Times as the main font
\usepackage{tabu}                      % only used for the table example
\usepackage{booktabs}                  % only used for the table example
\usepackage{lipsum}                    % used to generate placeholder text
\usepackage{mwe}                       % used to generate placeholder figures

\usepackage{mathptmx}                  % use matching math font

\usepackage{enumitem}

\onlineid{0}

\vgtccategory{Research}

\vgtcinsertpkg

\PassOptionsToPackage{naturalnames}{hyperref}
\title{Engaging Data-Art: Conducting a Public Hands-On Workshop}

\author{Jonathan C. Roberts\thanks{e-mail: j.c.roberts@bangor.ac.uk}\\ %
        \scriptsize Bangor University%
        }

\abstract{Data-art blends visualisation, data science, and artistic expression. It allows people to transform information and data into exciting and interesting visual narratives.  Hosting a public data-art hands-on workshop enables participants to engage with data and learn fundamental visualisation techniques. However, being a public event, it presents a range of challenges. We outline our approach to organising and conducting a public workshop, that caters to a wide age range, from children to adults. We divide the tutorial into three sections, focusing on data, sketching skills and visualisation. We place emphasis on public engagement, and ensure that participants have fun while learning new skills.}
\begin{document}
\firstsection{Introduction}
\maketitle

% JCR ---- Adding to remove backref 
\renewcommand*{\backref}[1]{
  % default interface
  % #1: backref list
  %
  % We want to use the alternative interface,
  % therefore the definition is empty here.
}

Data-art combines data science, visualisation, and art, allowing individuals to create artistic pieces that tell compelling stories from the underlying data. The subject of data-art creates an engaging and captivating public activity where individuals can come to an event to learn about science and produce intriguing artistic works. These data-art public hands-on workshops can be enjoyable for both participants and instructors, as they bring together a diverse group of individuals interested in art and data, fostering the development of creative skills. 

But developing a workshop that is both engaging and interesting, can be challenging. Participants arrive with diverse skill levels, attend for various reasons, and have different motivations for being there. The class must be engaging while imparting new skills. Furthermore the act of learning new skills can itself be challenging and may discourage participants from fully enjoying the tutorial, and the diverse age range of participants brings a variety of skills and levels of understanding. Each age group comes with different experiences and familiarity with cultural references, such as memes and real-life examples. This diversity necessitates a flexible approach to ensure that the tutorial is accessible and engaging for everyone, regardless of their background or prior knowledge. Therefore, it is essential to strike a balance between skill development, the subject matter, and having fun. The workshop must build on existing skills and guide participants further along their learning journey. At the end of the session we wish to see participants empowered with new skills, and excited to learn more about the topic. 

We detail the planning and delivery of a three-hour public data-art workshop. Our approach was to present key ideas succinctly, followed by immediate practical application through tasks. We  advertised the tutorial as a ``hands-on'' workshop. At every stage, we aimed for participant engagement. Instead of subjecting them to lengthy lectures, we focus on involving them practically with the concepts. We have created a simple data-art strategy, a three-part mantra, that can help people to understand how to perform data art, and what to consider. Our mantra is ``map these .. by these ... in this way''.
This approach enables creative data representations that provoke thought and spark interest, making complex information more engaging and accessible to a wide audience.

We organise the paper into three sections: preparation, delivery, and reflection of the tutorial. These sections offer a practical set of tasks and guidelines that others can easily follow.

\section{Background}
This data-art workshop is part of a wider set of events.  The hands-on workshop (which we explain in this paper), data-art projects created by students as part of their degree coursework, and an exhibition with student work and an open invitation to university researchers and local artists to showcase their data-art in the exhibition. The exhibition thus serves as the end-point for the public workshop, providing participants with the opportunity to create pieces that will be showcased alongside other entries in the exhibition. This enables them to feel that their efforts will be rewarded. We also aimed for participants to feel confident in applying their newly acquired skills beyond the workshop. Therefore, we prepared cheat sheets and notes that they could take home with them. 

Regarding the university course, we teach data-art in our undergraduate Creative Visualisation unit. The unit of study is an optional module for third-year computer science Bachelor of Science students.  It represents 20 UK Higher Education credits, which requires an average of 200 hours of study, and represents a sixth of the final BSc degree study. As part of their assessment for this module, students select a dataset, analyse its content, seek creative inspiration from established artists, and employ the Five Design-Sheet method~\cite{RobertsHeadleandRitsos16} to design their solution. Using \href{http://processing.org}{processing.org}, they then craft their data-art pieces, which are showcased in a public exhibition concluding their course. While some of the material is similar, in the university course and the hands-on workshop, the goals and depth are different. With the Creative Visualisation university module, we can go into detail, teach concepts over many weeks, and teach in-depth programming skills. Whereas the workshop is a 3 hour session, people have given up their Sunday afternoon to be with us at the workshop, and they are there to network, learn and engage with the material. Therefore our taught content, and approach are different. Whichever session though, our education philosophy is similar, we use an activity based approach, explaining certain concepts before getting participants to try the idea themselves.

\section{Related Work}
Data art is a creative field that mixes data science, visualisation, and artistic expression. The aim is to create visual and engaging narratives underpinned by data. Whereas data-science emphasises accuracy, data-art focuses on emphasising an emotional response, aesthetic beauty and developing engaging narratives. Vi\'egas and Wattenberg~\cite{ViegasWattenberg2007} describe the topic as ``\textit{visualisations of data created by artists with the intention of producing art}''. 
The objectives of data-art differ from those of scientific or information visualisation, which emphasise accuracy and exploration to enhance comprehension of data through precise visual representations. With data art people aim to convey narratives, evoke emotional responses, and prioritise aesthetic presentation over quantitative value representation. Another difference is that data artists use a wide range of mediums to convey their narratives. For example, `dear data'~\cite{lupi2016dear} use sketching, dataQuilt~\cite{zhang2020dataquilt} focus on glyphs,
\"Angeslev\"a and Cooper~\cite{RossJussi2004} in the `last clock' incorporate edited videos , 
Paolo Cirio (\href{https://www.paolocirio.net}{paolocirio.net}) integrate photographs, while Nathalie Miebach (\href{https://www.nathaliemiebach.com}{nathaliemiebach.com}) utilise rope and paper to display weather data.

Researchers express data-art in a variety of ways, highlighting the topic's wide-ranging applicability and significance. Kosara\cite{kosara2007visualization} proposed an interpretation that situates data art on a spectrum ranging from practical visualisation designs, through informative art displays, to artistic visualisations and sublime representations, akin to Milgram’s Virtual Reality continuum~\cite{milgram1995augmented}. While Ramirez~\cite{ramirez2008information} differentiates \textit{functional} and \textit{aesthetic} information visualisation. Functional visualisations focus on statistical analysis and emphasise clear communication. Whereas aesthetic visualisations ``are concerned with visually attractive forms of representation design''. Andrea and Vande Moere~\cite{AndreaMoere2007} likewise focus on aesthetics, presenting a two-dimensional model of \textit{intrinsic} (``cognitively effective visual mapping'') to \textit{extrinsic} (to facilitate communication of meaning, underpinned by the data) and on the horizontal axis: \textit{direct} to \textit{interpretive}. Where information visualisation is considered as being \textit{direct} and \textit{intrinsic}, whereas visualisation art is \textit{extrinsic} and \textit{interpretative}. Li~\cite{Li2018DataArt} argues that ``big data presents creative potential for digital artists''. Data helps to underpin the narrative, and give a direction to the art. However, art practices, visualisation, and often production practices differ. Corby~\cite{Corby2008} explains that data-art production processes are ``also novel and generally bear little resemblance to approaches in the sciences'', continuing that the visual representation of data in the visual arts has been led by graphic artists, or ``by renaissance teams'' as collaborations between scientists and artists~\cite{Cox2006}. Visualisation is appearing in the art curriculum, across ages; not only to help understand art-based data such as museum footfall, people's interests, or explaining the topic better~\cite{Grodoski2018}, but also to help students explore the topic and learn to create their own data-art pictures~\cite{BertlingETAL2021}.

%Tufte~\cite{Tufte1997} advocates for simplicity by eliminating unnecessary visual elements that do not significantly contribute to understanding the data. He suggests to remove chartjunk, explaining ``graphical elegance is often found in simplicity of design and complexity of data''~\cite{Tufte2001}.

%--------

\section{Preparation}
In preparation for the workshop there were several significant actions that we needed to achieve. We named the workshop ``stories with data'' and first decided when it would take place. We chose to hold the workshop at the end of the Creative Visualisation course, and before the exhibition opening. This enabled us to curate the work from participants for exhibition. We decided on a Sunday early afternoon, to allow people to attend without needing to take time off work. Second we wanted a venue that was centrally located and could readily run workshop formats. Our cinema and innovation centre, \href{htto://pontio.co.uk}{Pontio}, was ideal as it is centrally located and accessible, and the workshop space is setup with pens, paper, circular tables, a large screen, and so forth. We visited the venue, and looked at the resources, realising that we needed to purchase extra fine-tip pens and cardboard for the participants' artwork. The workshop venue was the same building as the exhibition, and had 15 large display boards on wheels. During the workshop we displayed preprints of the art exhibition from the exhibition, to enable participants to study the submissions. We purchased the missing card and pens.

Third we wanted to advertise the event. We needed to write the advert bilingually in English and Welsh, and added the event description to the University's event page and sent emails to local organisations and art clubs. We promoted the event through the university's events page and sent advertisements to local art clubs, social media groups, and key contacts at several nearby universities. Fourth we required people to register for the event. We set up an online shop, charging people no money, but requiring them to register. We limited the number of tickets to 35, aiming to create a more interactive experience for participants. Additionally, the room's capacity for workshop mode was restricted to 50 people. We printed a sign-in sheet to register the participants. Fifth, we wished to record and take photographs at the event. So to abide with UK law we printed the declaration and displayed it at the front of the room. Photography/ Filming in Progress. 
``Please note that filming/photography is taking place at this event for promotional and archival purposes. The photographs and recordings made are likely to appear on our website.
If you would prefer not to be photographed please let the photographer know.
For further information contact [our name]''. Sixth, we ordered catering and refreshments for the day and printed the hand-outs. Because we were using the work for evaluation we put the work through our ethics board.

\section{Execution}
We planned the workshop to be a hands-on experience. Our emphasis was on storytelling rather than presenting a huge swath of PowerPoint slides. While we did decide to use PowerPoint for the presentation, we kept the text to a minimum and tried to go to the activities with the least amount of explanation and presentation slides. We welcomed everyone and explained the room setup, emergency exits and so forth. Explained that we will take photographs and asked us to inform us if anyone did not wish to be recorded. We introduced ourselves and briefly explained our goals, which was to have fun, sketch, be creative, and create output that we can display in alongside the exhibition. 

The exhibition attracted a diverse group of participants. We welcomed families with children, including two ten-year-olds, several fourteen-year-olds, multiple teenagers, ten students, and some adults, many of whom accompanied the children. About a third of the students had previously taken other visualisation courses, while the rest were friends of these students.  One retired attendee noted that she came to connect with people at the university and to learn more about data-art.  We split the workshop into three parts, and had a short comfort break in between each part.

\newcommand{\subfigimg}[3][,]{%
  \setbox1=\hbox{\includegraphics[#1]{#3}}% Store image in box
  \leavevmode\rlap{\usebox1}% Print image
  \rlap{\hspace*{10pt}\raisebox{\dimexpr\ht1-2\baselineskip}{#2}}% Print label
  \phantom{\usebox1}% Insert appropriate spcing
}

\begin{figure}
    \centering
\includegraphics[alt={Four data art examples. Top (a) showing a photograph of paper cars, with green, yellow and red stripes. Middle left (b) is the poppy field visualisation (from poppyfield.org). Bottom left (c) is a group of coloured faces, based on Chernoff faces, that present World health data. Middle right photograph of the Ceramic poppies at Caernarfon Castle commemorating World War I soldiers.},width=\columnwidth]{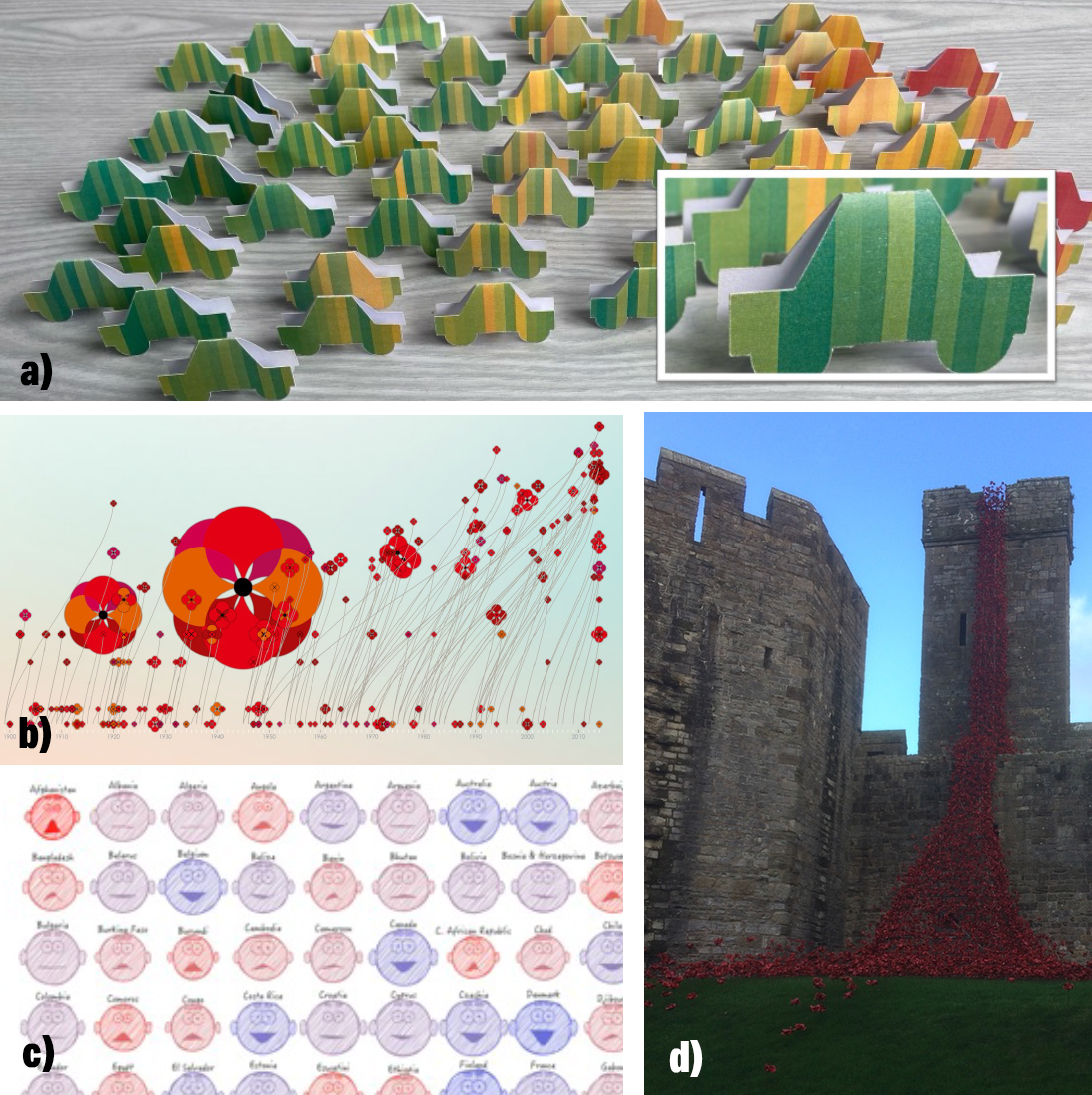}
    \caption{Four examples used at the start of the workshop. a) `student journey' showing physical paper cars of stripes of colour representative of their grades. b) `Poppy field' visualisation \href{https://www.poppyfield.org}{poppyfield.org} using poppies to present war fatalities. World health data represented by Chernoff faces\cite{Chernoff1973use}, artwork by author. Ceramic poppies in Caernarfon castle, UK. Poppies to commemorate each death in the British and Colonial forces of World War I; photograph by author. }
    \label{fig:openingstories}
\end{figure}

\subsection*{Part 1. What is data art?}
We explained data art as storytelling with data. Our goal is to engage people with data, and create expressive design solutions that excite, inform and even challenge people about the data and idea. Each data-art presents a story, that is created by encoding and utilising data in some way. In the workshop our emphasis is both to the visual design of the output, as well as how it tells a story. We do not wish to focus on accurate presentation of the data, instead focus on its appearance and how it can be used to engage viewers and participants with a narrative. To explain our point, we told four poignant stories, see~\cref{fig:openingstories}. 

First we showed `Student journeys'~\cite{mearman2019tangible}, \cref{fig:openingstories}a. We crafted paper cars to display student data, facilitating academic discussions about student progress. These cars serve as a ``talking stick'' --- the holder speaks and shares views about the student represented. The talking stick is a communication tool originating from indigenous cultures. It ensures orderly and respectful dialogue in group discussions. Only the person holding the stick (or in our case, a paper car) is allowed to speak, while others listen. This promotes active listening, equal participation, and thoughtful speech. The photograph, in  \cref{fig:openingstories}a, depicts the tangible artwork of forty paper cars, each symbolising a different student. The stripes on each car presents student grades over two study years. We asked participants to identify which cars represented high-performing students and which symbolised those performing less well. Cars with more red indicated students with lower performance, while those with more green stripes signified better performance. Participants found this mapping intuitive and expressed interest in learning more about the students themselves. They empathised with the students through the car metaphor.

Second we gave a live demonstration of the \href{https://www.poppyfield.org}{poppyfield.org} visualisation, \cref{fig:openingstories}b. It was designed by Valentina D’Efilippo and Nicolas Pigelet with data sourced primarily from The Polynational War Memorial. D’Efilippo and Pigelet, in their writings, emphasised that the project did not ``present an argument, nor is our intention to provide a conclusion on the topic'', going on to explain that they wish people ``to reflect on the human cost of war and to recognise the Great War Centenary''. 
Poppies are used to represent conflicts, with the stem indicating the war's duration, the flower size reflects the death toll, and colour variations denote regions involved. 
During the workshop, we prompted participants to describe what they observed and identify specific wars. One participant exclaimed, `\textit{`It's clear! The first and second world wars are represented by the two largest poppies}''. They were less certain about the height and placement of the poppy stems but grasped the meaning of the stem when it was explained to them. 

In the third example, we explained about the world happiness data survey~\cite{helliwell2024world}. We have developed a  Chernoff face inspired visualisation (see~\cref{fig:openingstories}c), that visualises this data in a cartoon style (using the handy processing.org library, \cite{Wood_etal2012}). We map the overall happiness value to both smile depth and facial colouration (red to blue). We map the value of `perception of social support' by smile width, generosity by brow length, GDP by overall face size, and life expectancy by ear length. We asked the workshop participants to identify the happiest and least happy countries. They enthusiastically shouted out their guesses. One participant, who had Pakistani heritage, was especially interested in Pakistan and its neighbouring countries.

Lastly, we discussed physical visualisations and presented images from the `Weeping Window' art installation  \cref{fig:openingstories}d. This installation featured thousands of ceramic poppies cascading down Caernarfon Castle in the UK. Each ceramic poppy honours a fallen soldier from the British and Colonial forces during World War I (see image credits). We described our visit to the installation as both visually stunning and deeply moving. Several participants recalled their experience, noting moments of silence in remembrance of the fallen soldiers. Data-driven art visualisations like these enable us to empathise with the data and evoke emotional responses.

\begin{figure}
    \centering
    \includegraphics[alt={A screenshot of the soundscapes tool, of where circles represent sound. It is a greyscale image that represents an x,y axis with time progressing along the x axis, and pitches on the y axis.},width=.99\columnwidth]{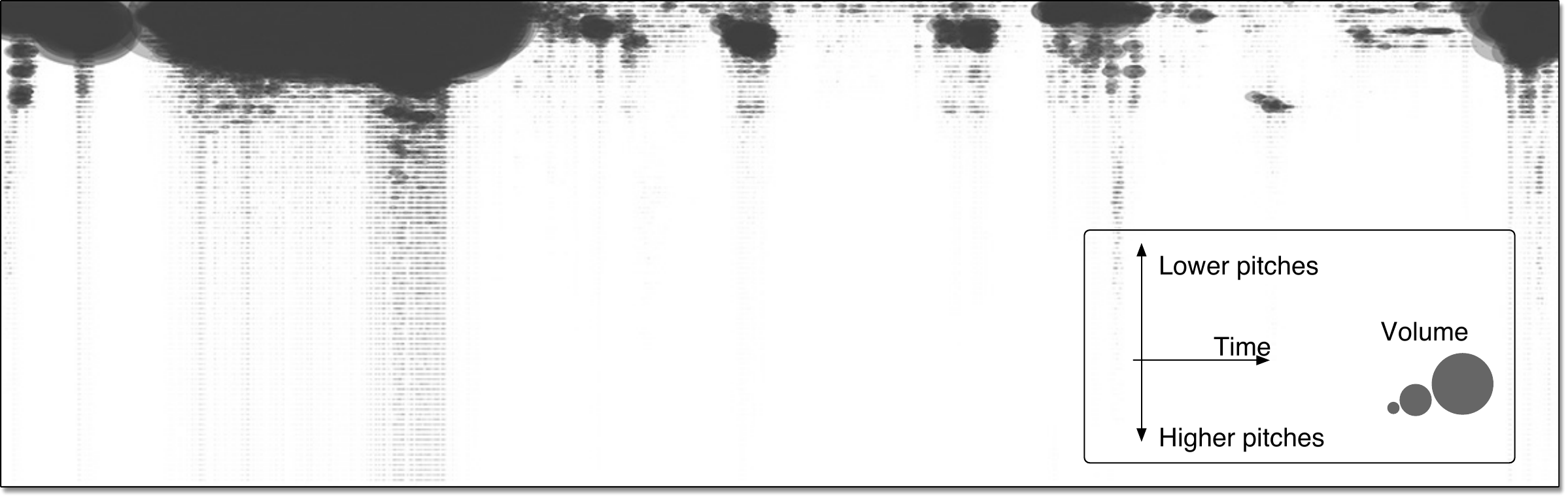}
    \caption{`Soundscapes' by author. Sound is turned into circles. Louder sounds are mapped to larger circles, higher pitches are lower down the screen, and time moves to the right. }
    \label{fig:soundscapes}
\end{figure}

After these examples, we presented the soundscapes visualisation (by author). This visualisation takes sound input, and changes it to a visual picture. Higher pitches are mapped to positions higher on the y-axis, while louder sounds corresponded to larger scaled objects. 
While running the software, we asked the participants to make some sounds—first by shouting ``hello'', then counting up to five in unison, and finally singing a note from a low pitch to a high pitch (\cref{fig:soundscapes}). On playback, we asked the participants to explain how the sound was mapped to the picture.

By showcasing these examples, we prompted participants to explore different data and mapping techniques. In the soundscapes demonstration, they grasped several mappings of loudness and pitch, but needed to be prompted to see that time moves to the right of the page. In the poppy field illustration, they comprehended the key narrative elements of the major wars. In the paper car scenario, they quickly understood the mappings of grades for both high-achieving and lower-achieving students. We followed this discussion with some examples of traditional visualisation mappings, and explaining the retinal variables (size, length, shape, orientation, colour, texture, transparency, etc. \cite{Bertin1983}).

\begin{figure}
    \centering
    \includegraphics[alt={Several sketches of hands. One large one on the left with text to explain the categories. Eleven smaller sketches on the right, from workshop participants.},width=\columnwidth]{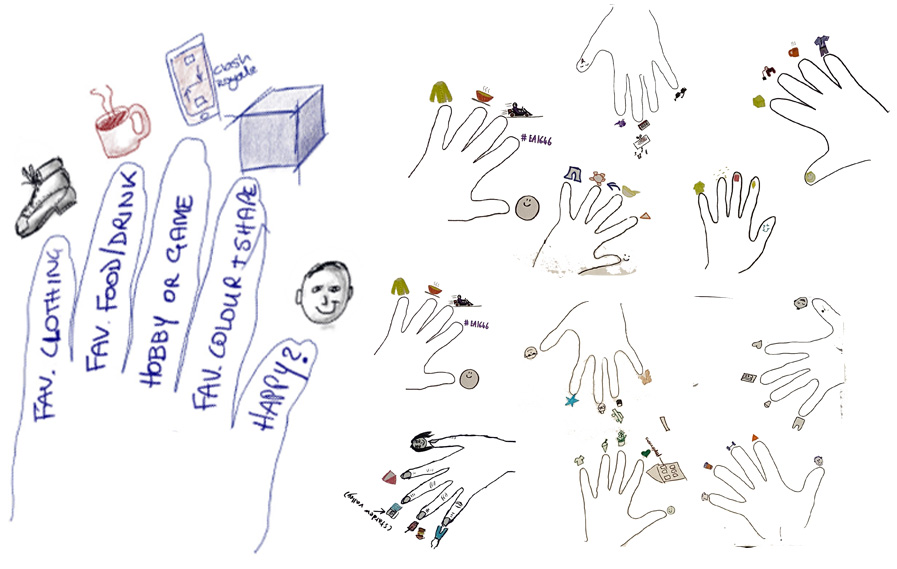}
    \caption{The sketch round your hand activity, where participants add five sketches, one to each finger: smiley face, favourite shape and colour, hobby or game they play, favourite food or drink, and clothing or accessory.}
    \label{fig:hands}
\end{figure}
In the final activity, we asked people to draw round their hands, \cref{fig:hands}. We asked them to sketch a smiley face on the thumb (or however they were feeling), their favourite shape with their favourite colour on the index finger, hobby or game on their middle, favourite food or drink on their ring finger, and finally favourite clothing or accessory on their pinky. We examined the various shapes and sizes of hands, which represent different ages. We discussed how each person interpreted the activity and what the items symbolised for them personally. As a result, participants began to share their own stories, and start to imagine how they can map data to pictures and create their own artwork.

\subsection*{Part 2. Sketching skills, preparing for data-sketching}
The plan for the second half was to focus on sketching skills. We described ten activities, to help people develop key skills. From drawing lines, thinking about shading and colours and adding in effects. These sketching skills help people to prepare for people to perform data-sketching. To facilitate recalling each part of the sketching activity, we named them and included notes describing each task as follows:
\setlist[itemize]{leftmargin=*}
\begin{itemize}[itemsep=-1mm]
    \item \textbf{Look}: Focus on the destination of the line when drawing. This helps ensure your lines are directed and purposeful, improving accuracy and control.
    \item \textbf{Arm}: Use your arm or elbow instead of your wrist when drawing lines and circles. This technique provides more stability and allows for smoother, more consistent strokes.
    \item \textbf{Move Canvas}: Move the canvas, not your hand, especially when drawing long lines. Shifting the canvas helps maintain a consistent angle and reduces the strain on your wrist.
    \item \textbf{Natural Arc}: Use the natural arc of your hand or arm to draw curves. This method leverages your body's natural movement, making curves more fluid and natural-looking.
    \item \textbf{Few Colours}: Use a limited color palette (2 or 3 analogous colors plus 1 highlight). This approach keeps your artwork cohesive and visually appealing without overwhelming the viewer.
    \item \textbf{Where’s the Sun?}: Always consider the light source when shading. Knowing the imaginary position of the sun helps create realistic shadows and highlights, adding depth to your drawing.
    \item \textbf{Simple Shading}: Avoid filling everything with color. Instead, use shading techniques over lines to add dimension and texture without making the artwork look flat.
    \item \textbf{Visualise Ahead}: Plan where elements will go before starting to draw. Think about the composition and the emotions you want to convey, which helps create a more coherent and impactful piece.
    \item \textbf{People Have Feelings!}: When drawing people, give them expressions and body language that reflect emotions. This makes your characters more relatable and engaging.
    \item \textbf{Add Effects}: Incorporate a few well-placed effects, such as lines for emphasis, to enhance the visual impact of your artwork. Use these sparingly to avoid cluttering the piece.
\end{itemize}

\begin{figure}
    \centering
    \includegraphics[alt={Two data-art images. The top image features a set of thumbnails on the left with lines extending from each. The bottom image shows book cover thumbnails along the bottom, with text emerging from each thumbnail.},width=\columnwidth]{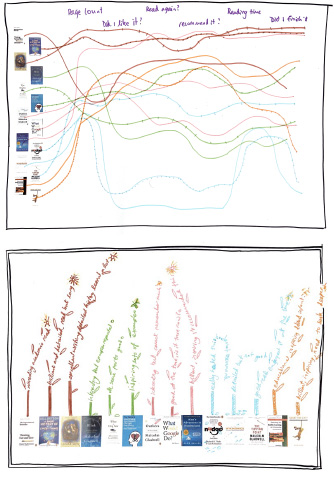}
    \caption{Two distinct visualisations depict fourteen books, illustrating whether the book was enjoyed, completed, recommended, and other related factors. The top visualisation is inspired by the parallel coordinates plot (PCP) technique \cite{inselberg1990parallel}, while the bottom visualisation presents a brief sentence indicating enjoyment -- represented by an upward line if the book was enjoyed, or a drooping plant stem if it was not. }
    \label{fig:booksVis}
\end{figure}
\subsection*{Part 3. }
The third part draws the ideas together. We explain the three-part mantra: ``map these .. by these .. in this way''. The process begins by initially focusing on the data variables and the narrative they convey. Next, determine which visual (retinal) variables—such as length, size, and shape—should be employed to effectively present the data. Lastly, carefully consider the arrangement and organisation of the artwork, including spatial relationships and overall visual composition. This involves deciding whether elements are grouped closely or spaced apart, and how the appearance of the display contributes to its meaning.
To get participants to understand the thought process we gave a live demonstration. We described fourteen books that had been read; some we enjoyed, others we struggled to complete, while others were not completed.  We explored two visualisation designs. The first (inspired by a parallel coordinate plot visualisation~\cite{inselberg1990parallel}) featured lines that ascended for higher values and descended for lower values across the page, see~\cref{fig:booksVis}. 

\begin{figure}
    \centering
    \includegraphics[alt={Three data-art images, each displaying a thumbnail of a different food item on the left, with lines extending to the right from each thumbnail.},width=\columnwidth]{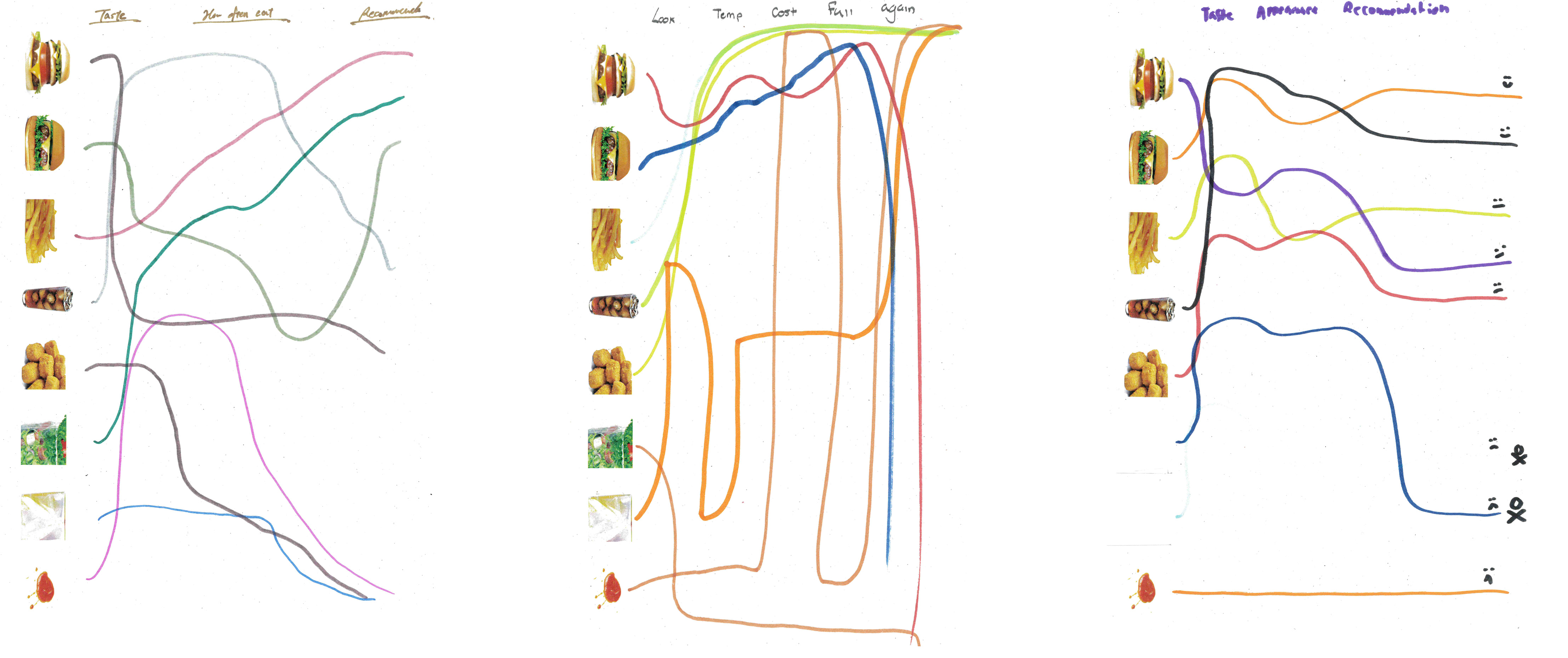}
    \caption{Participants translated the parallel plot design into action during the ``Food Critic'' activity, where they evaluated images of various burgers, fries, and other fast-food items commonly found at fast-food outlets. One participant (shown in the right sketch) strongly disliked the apple slices and salad, going so far as to cover the image with a white scrap of paper.}
    \label{fig:foodCritic}
\end{figure}

The second data-art sketch showcases a straightforward sentence that indicates whether the book was enjoyable and worthy of recommendation. If the book was difficult to read, the words appeared in a drooping manner.
Participants took these ideas and applied them to consider fast-food. In an activity that we name `food critic', we display images of different burgers, fries and food items that may be purchased at a fast-food outlet. \Cref{fig:foodCritic} shows the picture from three participants. Some participants did not like the healthy options, choosing to remove the back of apple slices, or the salad. They were telling their own story, crafting a display, that was inspired from data.  
After a break we showed many examples of data art, including examples that visualised mathematics formula, transport data, internet traffic, and musicians over time. Next we ran a longer activity, focused on applying the ideas, and getting participants to be more inventive with their ideas. 

\begin{figure}
    \centering
    \includegraphics[alt={Three sketches. The left (a) depicts a large tree with a skyscraper-like structure. The top right (b) shows a typical house sketch with a tree on the right. The bottom right (c) features a purple sketch of a house with a tree on the left.},width=\columnwidth]{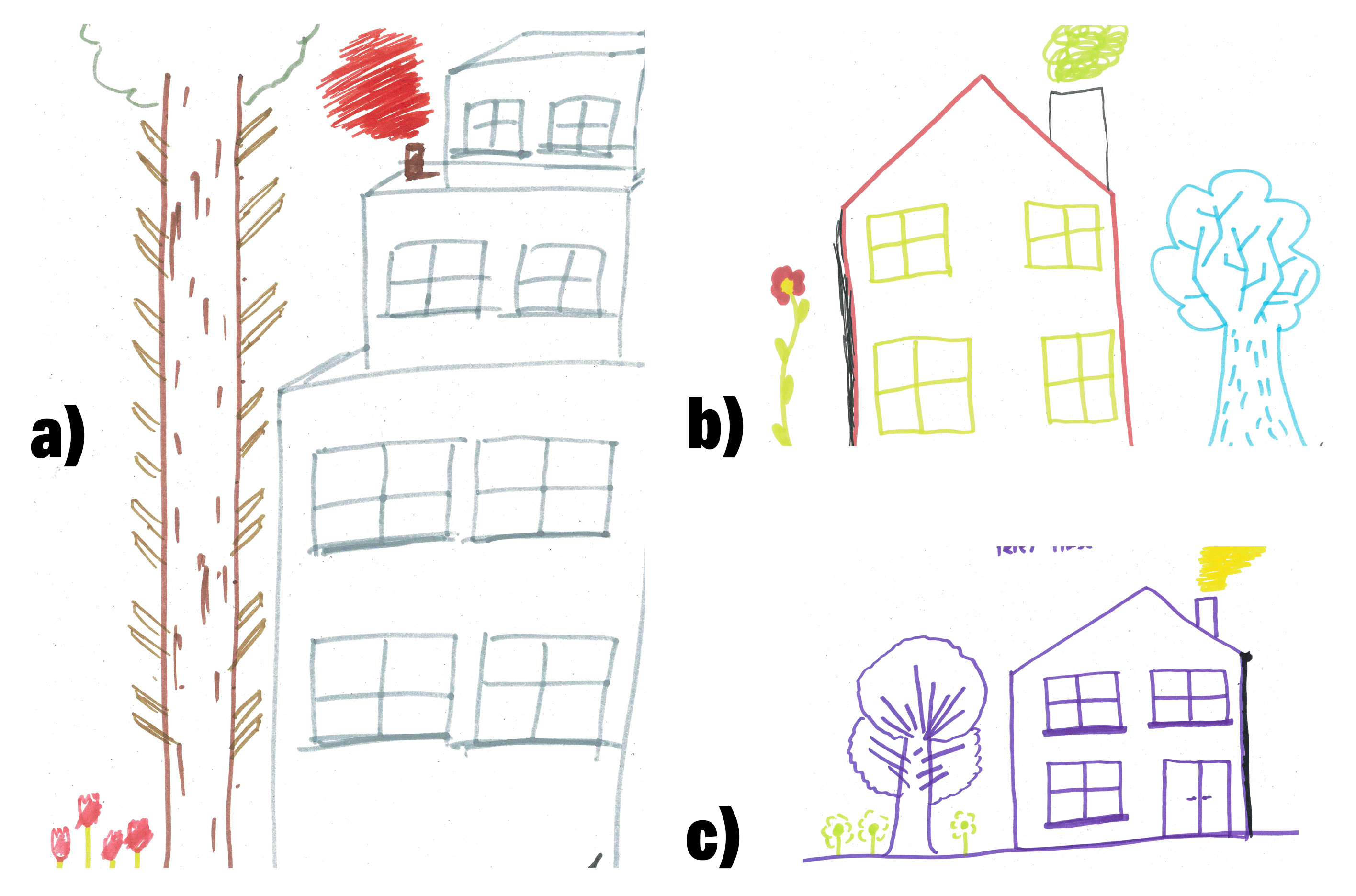}
    \caption{Three participants' results of the `my house' activity. The first (a) lived in university housing with many people, the second and third lived with one other person. They mapped people to aspects of the tree, and computers and technology to windows.}
    \label{fig:myHouse}
\end{figure}
In this `my house' activity they were asked to consider where they live, and sketch a creative picture. Different aspects of the picture should correspond to quantities. For example, the quantity of cars in the household could be mapped to the quantity of pot plants. The number of people in the house is represented by the quantity of windows. The quantity of animals in the house, could be mapped to trees by the side of the house.  They were tasked in groups to first decide on their mappings. They needed to write down how they would ``map these .. by these .. in this way''. Each group member would then take these mappings and describe their situation in this way.

\section{Discussion and Conclusion}
Throughout the workshop, emphasis was placed on the narrative potential of data visualisation. The idea behind each visual creation was not merely as a representation of data points, but as a medium for crafting compelling stories. Whether illustrating consumer preferences or market trends in the food critic activity, or looking at where they lived in the~my house' activity, each aimed to get participants to discuss the narrative of the data. We had positive responses from our feedback questionnaire. One participant wrote ``\textit{I really liked the hands-on activities. It was really engaging}'', another wrote ``\textit{very informative and fun, felt easy to understand}''. This was positive feedback, as we were hoping to create a set of activities that people engaged with, and learn from.Another person wrote ``\textit{Fun, engaging. Very helpful for people studying data art next year}''. 

When asked about negatives or dislikes, we also received some positive feedback. One wrote ``\textit{more activities}'' another wanted ``\textit{pencils / other mediums even paint}''. On this matter, we did consider other mediums, such as paint, cardboard cutting, but decided that it would need to be a longer session to add in other media. Also, paint can be very messy and difficult to control which could help to add additional narrative elements to the works, but also may distract from the final art piece. One participant wrote that the ``\textit{instructions could be clearer for people who may not understand}''. We were deliberately vague at some points, because we wanted the participants to be inventive and creative, and did not want to lead. 
We also received some positive comments on the sketching part of the workshop. One participant afterwards said verbally that ``\textit{they now know how to draw a straight line (through looking at the end point) and a circle (with their elbow and arm)}'' and that ``\textit{it had changed the way they draw}''.

We enjoyed running the workshop. On reflection we feel that there are several powerful lessons that we have learnt. The strategy of swiftly getting to the hands-on activities proved successful. It was gratifying to observe participants improve their skills and engage in creative thinking. Also, we felt that because we had advertised it as a hands-on workshop,  that the participants should be the ones doing the work. The art skills proved highly effective, allowing for immediate examples and participant application. The ten lessons appeared fitting for the session. While additional lessons could have been included, they may be reserved for a future, extended workshop. 
Additionally, naming the activities simplified referencing and reflection. Each lesson was crafted to reinforce a concept and aid participant learning. Thus, referring back to earlier activities was crucial. It also enabled participants to articulate their preferences and critiques afterwards. One participant remarked upon leaving that they particularly enjoyed the food-critic activity, humorously commenting, ``\textit{who goes to a fast-food outlet and eats a salad}?'' 
Presenting numerous examples of artists' work proved to be impactful. It was crucial to showcase a diverse range, spanning from intricately crafted cars to the poignant depiction of war casualties in the poppy visualisation, and even lighthearted and humorous Chernoff face inspired artwork. 

For the exhibition, we decided to showcase approximately half of the workshop's output. We chose these submissions, because while most participants completed the tasks, some were notably clearer and more polished. We set up two large poster boards alongside the exhibit to display the output from the workshop. We included photographs from the workshop, explanations of some tasks, and example artwork.  Consequently, the exhibition featured three sections: students' work from the Creative Visualisation course, research and artist entries, and the workshop output. This approach provided a diverse range of completed work, from meticulously crafted pieces to quicker visualisation styles.

Several practical aspects of leading and running a workshop must be considered. These include deciding on and purchasing resources, arranging refreshments, securing the venue, organising the booking system, addressing legal issues, and more. Additionally, it was essential to schedule breaks throughout the three-hour session. Each segment lasted about 50 minutes, followed by a ten-minute break. Organising workshops is tiring for both the organiser and the participants, so ensuring adequate rest, refreshments, and breaks is crucial.

In conclusion, participants enjoyed themselves and seemed to acquire new skills. The feedback was positive, and encouraging of running further workshops. They saw the passion for data and art from us, with one participant saying ``\textit{can see your passion}''. Participants provided several suggestions for other workshops, including  ``\textit{different media}'', ``\textit{digital}'', ``\textit{maths / writing stories}'', ``\textit{maths or science}'', and ``\textit{another data-art workshop}''. One participant also expressed a request of expert lessons in data management, information visualisation and art, such as portrait drawing. There was a clear appetite for more, and it was gratifying to see such enthusiasm.

\section*{Figure Credits}
\label{sec:figure_credits}
\cref{fig:openingstories}. a) Photograph of `student journey' cars by author. b) `Poppy field' visualisation credit to \href{https://www.poppyfield.org}{poppyfield.org}. c) Chernoff faces inspired artwork, credit author. d) Ceramic poppies in Caernarfon castle, UK, photograph by author. Conceived by artist Paul Cummins and designed by Tom Piper, the poppies were crafted by Paul Cummins Ceramics Limited in collaboration with Historic Royal Palaces. \Cref{fig:soundscapes,fig:booksVis} the author, and \cref{fig:foodCritic,fig:myHouse} the author, from art submitted to the workshop.

\bibliographystyle{abbrv-doi-hyperref-narrow}
\bibliography{data-art-workshop-eduVis2024.bib}     

\end{document}